\documentclass[twocolumn,amssymb, amsmath,
nofootinbib,tightenlines,nobibnotes,aps,prb]{revtex4}
\usepackage[dvips]{graphicx}
\usepackage[T1]{fontenc}
\usepackage{dcolumn}
\usepackage{bm}
\usepackage{natbib}
\usepackage{amssymb}
\usepackage{color}
\usepackage{dcolumn}
\newcommand{\figjov}[1]{\includegraphics[width=0.45\textwidth]{#1.EPS}}

\begin{document}

\title{Anomalous magnetoresistance behavior of CoFe nano-oxide spin valves at
low temperatures}

\author{J. Ventura}\email{joventur@fc.up.pt}
\author{J. B. Sousa}
\author{M. A. Salgueiro da Silva}
\affiliation{IFIMUP and Faculty of Sciences U. Porto, Rua do Campo
Alegre, 678, 4169-007, Porto, Portugal}

\author{P. P. Freitas and A. Veloso}
\affiliation{INESC, Solid State Technical Group, R. Alves Redol,
9-1, Lisbon, 1000, Portugal}

\begin{abstract}
We report magnetoresistance curves of CoFe nano-oxide specular spin
valves of MnIr/CoFe/nano-oxidized CoFe/CoFe/Cu/CoFe/nano-oxidized
CoFe/Ta at different temperatures from 300 to 20 K. We extend the
Stoner-Wolfarth model of a common spin valve to a specular spin
valve, introducing the separation of the pinned layer into two
sublayers and their magnetic coupling across the nano-oxide. We
study the effect of different coupling/exchange (between the
antiferromagnetic layer and the bottom sublayer) field ratios on the
magnetization and magnetoresistance, corresponding with the
experimentally observed anomalous bumps in low temperature
magnetoresistance curves.
\end{abstract}
\maketitle

\section{INTRODUCTION}
A simple spin valve\cite{Dieny} (SV) is a nanostructure with two
ferromagnetic (FM) layers separated by a sufficiently thick
nonmagnetic (NM) spacer. An adjacent antiferromagnetic (AFM)
material fixes the magnetization of one of the FM layers, the
so-called pinned layer. The other FM layer, called the free layer,
is only weakly coupled (magnetically) to the pinned layer. Spin up
and spin down electrons are differently scattered (either in the
bulk or at the interfaces), usually producing a large
magnetoresistance (MR) when a small applied magnetic field H
reverses the free layer magnetization with respect to that of the
pinned layer.

Recent reports on spin valves with the pinned and/or free layer
partially
oxidized,\cite{Veloso(especular)6,Gillies3,Sakakima(especular)4}
showed great MR enhancement over the conventional (nonoxidized) spin
valves (CSV). Electrons are believed to reflect specularly at the
nano-oxide layer (NOL)/FM interfaces, thus yielding higher MR
ratios. Low temperature and temperature dependence studies of the
transport and magnetic properties of NOL SVs can provide valuable
information on the oxide layer properties, also assisting in the
optimization of NOL SVs and corresponding physical
understanding.\cite{Gillies3,Sousa(comparative)5}

\begin{figure}
\begin{center}
\figjov{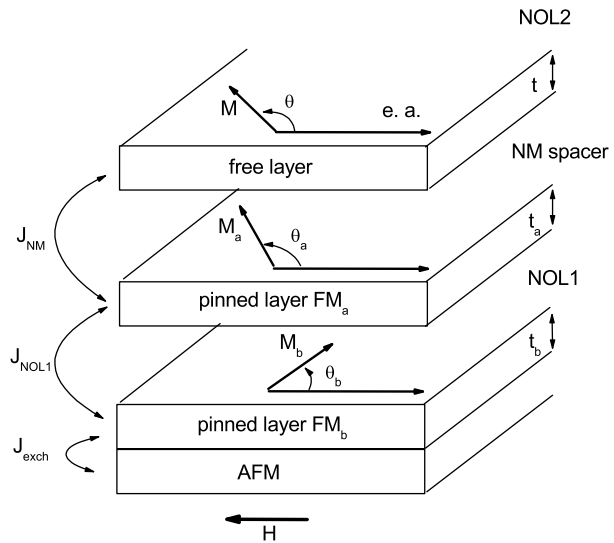} \caption{Structure of a NOL SV.
\label{fig:NOL_SV}}
\end{center}
\end{figure}

We previously reported\cite{Sousa(comparative)5} the temperature
dependence of the magnetoresistance MR(H) curves for NOL SVs with
the structure seed/MnIr/CoFe/oxidation (NOL1)/CoFe/Cu/CoFe/
oxidation (NOL2). The first FM deposited layer (CoFe on MnIr) will
be called the below-NOL1 pinned layer (FM$_b$) and its upper part is
oxidized over an adequate thickness to form the NOL1 layer. The FM
layer deposited after this oxidation will be called the above-NOL1
pinned layer (FM$_a$). The pinned layer thus consists of both FM$_b$
and FM$_a$ sublayers, separated by the NOL1 oxide layer. Our
study\cite{Sousa(comparative)5} included the temperature dependence
of the M(H) and MR(H) curves for such NOL SVs, revealing the
appearance of several anomalous features at low temperature, namely
anomalous bumps in MR(H) at intermediate fields and the absence of
complete MR saturation up to large positive fields. In this article
we focus on the physical origin of such features. For this we
present a model based on the total energy\cite{Parker6} of a NOL SV
to describe the magnetization orientation in each of the three FM
layers under an external magnetic field and the resulting MR(H)
behavior. We assume that the FM$_b$ and FM$_a$ layers are
ferromagnetically coupled across NOL1 (Fig.
\ref{fig:NOL_SV}),\cite{Gillies3} and study the effect of such
coupling strength on the magnetization M(H) and MR(H) curves. The
model accounts for the observed anomalous MR(H)
bumps,\cite{Sousa(comparative)5} relating them to the M reversal in
the pinned layer.

The absence of full alignment in large positive fields at low
temperature is discussed in terms of the permanence of 360$^\circ$
domain walls; the underlying H asymmetry is related to the dc field
anneal. The key role played by the nano-oxide CoFe specular layers
is also emphasized.

\section{THEORETICAL MODEL}
We extend the simple model describing the total energy of a
CSV\cite{Parker6} to the more complex structure of NOL SVs,
explicitly introducing the pinned layer partition into the FM$_b$
and FM$_a$ sublayers separated by NOL1. The exchange coupling
between the FM$_b$ and AFM layers fixes the easy axis of the
magnetization in the FM$_b$ layer and (as a consequence of coupling
interactions) in the FM$_a$ and free layers. We assume coherent
rotation of the magnetization within each layer, under an external
magnetic field H. To simplify the treatment no anisotropy terms are
included. The total energy per unit area $E$ of a NOL SV is thus
written as the sum of Zeeman E$_{Zeeman}$, coupling E$_{coup}$, and
exchange E$_{exch}$ energies (H along the easy axis):
\begin{equation}
\begin{split}
E_{Zeeman}=-\mu_{0}M_{b}t_{b}H\cos\theta_{b}&-\mu_{0}M_{a}t_{a}H\cos\theta_{a}\\
&-\mu_{0}MtH\cos\theta,
\end{split}
\end{equation}
\begin{equation}
E_{coup}=-J_{NOL1}\cos(\theta_{b}-\theta_{a})-J_{NM}\cos(\theta_{a}-\theta),
\end{equation}
\begin{equation}
E_{exch}=-J_{exch}\cos\theta_{b}.
\end{equation}
M$_b$ ($\theta_b$), M$_a$ ($\theta_a$), and M ($\theta$) are the
saturation magnetizations (angles of the magnetization with the easy
axis) in the FM$_b$, FM$_a$, and free layers, respectively.
J$_{NOL1}$ (J$_{NM}$) is the ferromagnetic exchange coupling energy
per unit area between the FM$_b$ and FM$_a$ (FM$_a$ and free FM)
layers and J$_{exch}$ is the exchange bias energy per unit area
between the AFM and FM$_b$ layers. Here $t_b$, $t_a$, and $t$ are
the thicknesses of the FM$_b$ , FM$_a$ (assumed equal), and free
layers (thicker than the previous), respectively.

\begin{figure}
\begin{center}
\figjov{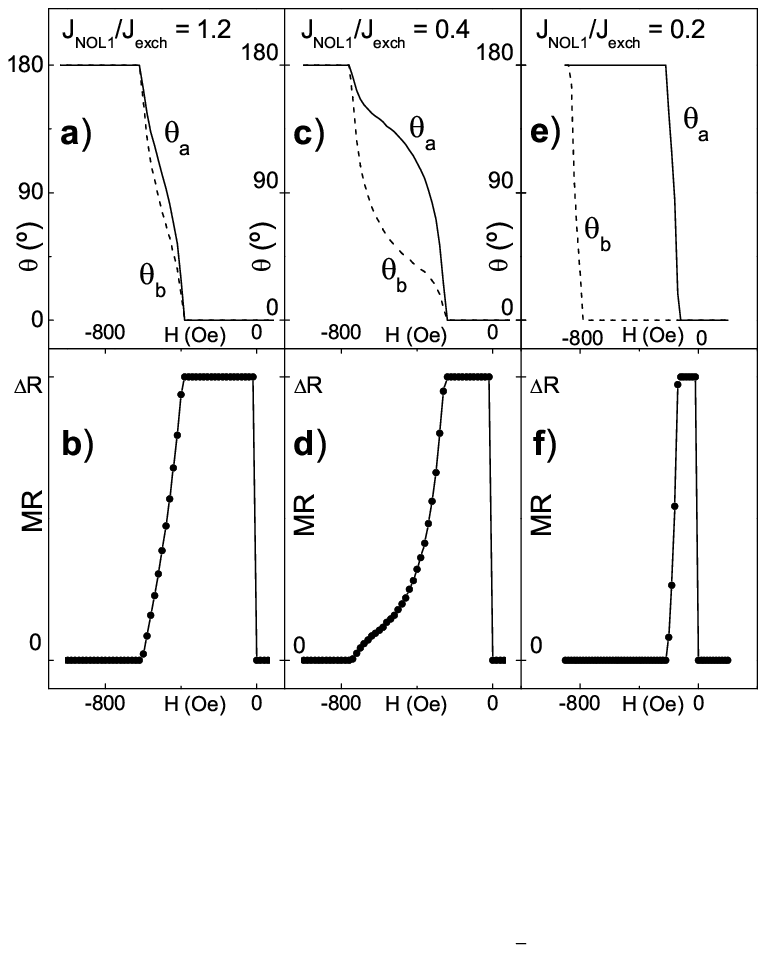} \caption{Model simulations using different J$_{exch}$
/J$_{NOL1}$ ratios.\label{fig:MReM}}\end{center}
\end{figure}

We numerically obtain the $\theta_b$, $\theta_a$, and $\theta$
angles which minimize E for each value of H and then calculate the
total magnetic moment per unit area $m$ of the NOL SV appearing
along the easy axis:
\begin{equation}
\begin{split}
m(H)=M_{b}t_{b}\cos(\theta_{b}(H))&+M_{a}t_{a}\cos(\theta_{a}(H))\\
&+Mt\cos(\theta(H)). \label{eq:m_H}
\end{split}
\end{equation}
Dieny et al.\cite{Dieny-JAP7} showed that in a CoFe/Cu/SyAP/MnPt
spin valve (SyAP: synthetic AFM\cite{VelosoeFreitas8} pinned layer
of CoFe/Ru/CoFe), the magnetoresistance associated with the two CoFe
sublayers separated by Ru is only a small fraction of total MR. As a
first approximation we then assume that MR is only due to the
relative orientation of the magnetization in the FM$_a$ and free
layer:
\begin{equation}
MR(H)=\Delta
R\left(\frac{1-\cos\left[\theta_{a}(H)-\theta(H)\right]}{2}\right),\label{eq:MR}\end{equation}
where $\Delta R$ is a characteristic amplitude. As expected for very
small J$_{NM}$ (across the spacer) the almost uncoupled free layer
always reverses suddenly at very low negative values of H, but the
model predicts three regimes for the rotation of the pinned layer at
higher fields, depending on the J$_{NOL1}$/J$_{exch}$ ratio:

(i) J$_{NOL1}>$J$_{exch}$  [Fig. \ref{fig:MReM}(a)]: the
magnetization of both FM$_b$ and FM$_a$ sublayers fully rotate
($0\rightarrow\pi$) in a narrow field interval and always with
approximately the same angle. This occurs because the large magnetic
coupling between FM$_b$ and FM$_a$ (compared to J$_{exch}$) forces
the two layers to behave as a (single) intra-pinned layer. Figure
\ref{fig:MReM}(b) shows the predicted MR(H) curve, displaying an
abrupt MR drop when the magnetizations of FM$_b$ and FM$_a$ rotate.

(ii) J$_{NOL1}\leqslant$J$_{exch}$ [Fig. \ref{fig:MReM}(c)]: both
FM$_b$ and FM$_a$ layers reverse in the same field interval, but the
corresponding range increases with decreasing J$_{NOL1}$ and a
considerable dephasing occurs between $\theta_b$ and $\theta_a$ at
intermediate fields. Such increase in ($\theta_b-\theta_a$) leads to
broadening in the M(H) and MR(H) curves. Figure \ref{fig:MReM}(d)
illustrates, for J$_{NOL1}=0.4$~J$_{exch}$ , the predicted bump in
the MR(H) curve at sufficiently negative fields.

(iii) J$_{NOL1}\ll$J$_{exch}$ [Fig. \ref{fig:MReM}(e)]: the
magnetizations of FM$_b$ and FM$_a$ now reverse almost individually
and over distinct narrow field ranges; a small disturbance in the
other's magnetization angle occurs in such ranges, producing also a
disturbance in MR [Fig. \ref{fig:MReM}(f)]. If J$_{NOL1}$ is further
decreased, both M$_b$ and M$_a$ reversals become truly independent,
i.e., no change in the angle of the magnetization of one layer is
visible when the other rotates. This produces three steps in the
M(H) curve, corresponding to the magnetization reversal in the
FM$_b$, FM$_a$, and free FM layers. No variation occurs in MR due to
FM$_b$ magnetization reversal [Eq. \ref{eq:MR}].

\section{EXPERIMENTAL MR(T;H) BEHAVIOR}
The details on the spin valves preparation and measurement
techniques have been previously reported\cite{Sousa(comparative)5}
and will not be presented here. We simply comment that after being
post-annealed in vacuum (10$^-6$ Torr) at 270$^\circ$C for 10 min
the spin valves were cooled in a 3 kOe applied field, to impress an
easy magnetic axis.

\begin{figure}
\begin{center}
\figjov{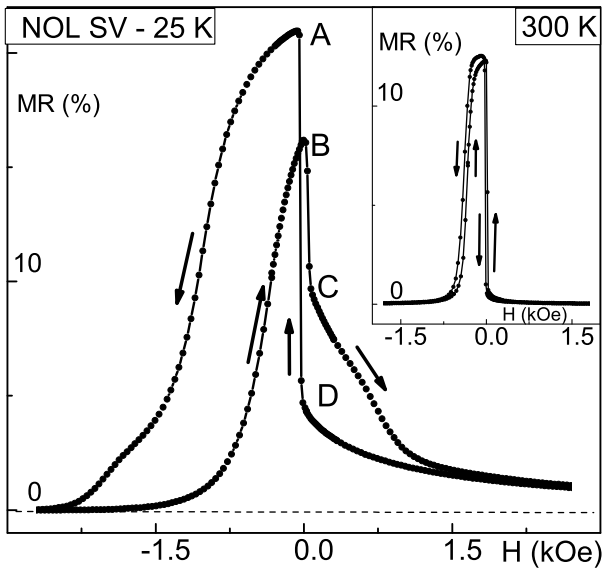}
\caption{Magnetoresistance of NOL SV at low (25 K)
and high (300 K) temperature.\label{MR_NOL}}\end{center}
\end{figure}

Room temperature MR(H) curves for a NOL SV with the structure Ta(67
Å)/Ni$_{81}$Fe$_{19}$(42 Å)/Mn$_{83}$Ir$_{17}$(90 Å)/CoFe(14
Å)/oxidation/Co$_{90}$Fe$_{10}$(15 Å)/Cu(22 Å)/Co$_{90}$Fe$_{10}$(40
Å)/oxidation/Ta(30 Å) exhibit the usual SV behavior (inset of Fig.
\ref{MR_NOL}): in a positive field both pinned and free layers are
parallel aligned ($\uparrow\uparrow$), but the magnetization of the
free layer abruptly reverses in a small negative field
($\downarrow\uparrow$ alignment results), leading to a high
resistance over a finite $\Delta H$ range. Parallel alignment in the
opposite sense ($\downarrow\downarrow$) ultimately occurs when the
negative field overcomes the exchange bias between the AFM and the
pinned layer, leading again to the low resistance state. The
incorporation of the NOL greatly enhances the MR ratio over that
observed in the corresponding nonspecular SV, from 5.9\% to 12.5\%
at room temperature in our case. This ratio was found to increase
linearly with decreasing temperature,\cite{Sousa(comparative)5} due
to the decrease in electron-spin wave scattering in the FM
layers.\cite{Chaiken9}

With decreasing temperature the following features arise in the
MR(H) curves of the NOL SV (Fig.
\ref{MR_NOL}):\cite{Sousa(comparative)5} Large MR(H) broadening and
the emergence of two anomalous MR bumps, one at each side;
incomplete $\uparrow\uparrow$ alignment in positive fields up to
H$_{max}=8$ kOe (MR does not vanish), whereas full
$\downarrow\downarrow$ alignment (MR=0) occurs under moderate
negative fields; absence of the usual $\Delta H$ plateau of constant
(maximum) MR at low temperatures, indicating imperfect magnetization
antiparallelism; and large hysteretic MR(H) cycles. These features
are related to the presence of the NOL, since they are absent (or
rather small) in the CSV.\cite{Sousa(comparative)5}

\section{DISCUSSION AND CONCLUSIONS}
Increasing the applied field from negative H$_{max}$ , where
complete parallelism $\downarrow\downarrow$ occurs, initiates the M
reversal of the pinned layer (exchange coupling with MnIr favors
positive magnetization) producing a gradual MR increase. However,
when the field reaches small positive values the free layer suddenly
reverses its magnetization, producing a discontinuous MR decrease.
One notices that the reversal of the pinned layer is still not
complete since MR does not reach the value observed in the
decreasing field regime (see Fig. \ref{MR_NOL} and below). To
confirm that the rotation of the pinned layer continues under
positive fields, we plot the difference between the maximum MR for
decreasing and increasing fields (points A and B in Fig.
\ref{MR_NOL}) and the difference between after-free-layer-reversal
in increasing and before-free-layer-reversal in decreasing fields
(points C and D in the same figure) as a function of temperature
(not shown). These differences are indeed similar, showing that the
anomalous MR bump in positive fields is due to the incomplete
reversal of the pinned layer. A rise of (both) these differences is
visible below $\sim200$ K, which may be related to the presence of
an AFM oxide in the NOLs.

In MnIr/CoFe bilayers it was shown that magnetization reversal
proceeds by coherent rotation, nucleation, and motion of domain
walls.\cite{Wang_MnIr_rotation} We believe that stable domain walls
still remain in the pinned layer under positive fields and only
gradually disappear as H increases, thus preventing full parallelism
of the layer magnetic moments. In fact, at low temperature we were
unable to achieve zero MR even at positive H$_{max}=$8 kOe. It was
recently found in CoFe thin films that 360$^\circ$ domain walls can
indeed be stable up to high fields (rather than annihilating) and
can give a significant contribution to the electrical
resistance.\cite{Tiusan9}

For decreasing fields from H$_{max}$ , the free layer magnetization
suddenly reverses at a small negative field but maximum MR (full
magnetic antiparallelism FM$_a$/free layer) cannot be achieved due
to the previous partial rotation/domain walls in the pinned layer;
no $\Delta H$ plateau of maximum MR should then occur. Further field
decrease keeps the ongoing rotation of the pinned layer and produces
a broad decrease of the MR ratio due to the FM$_b$/FM$_a$ coupling
(see model in Sec. II). When the field value reaches the bump in the
left side of the MR(H) curve, the rotation of the FM$_a$
magnetization is almost complete, but FM$_b$ is still far from
complete rotation [Figs. 2(c)-2(d)], due to the higher exchange
pinning with the AFM. Complete parallel alignment is only obtained
at a higher (negative) field. The model presented in Sec. II
describes well important characteristics of this descending branch
of the MR(H) curve (broadening and MR bump) for J$_{NOL1}\approx$0.4
J$_{exch}$ . Because these anomalous features appear and grow with
decreasing temperature, one concludes that J$_{NOL1}$ grows at a
slower rate with decreasing temperature than J$_{exch}$.

The model we presented here gives a fair description of the
anomalous bump present in the MR(H) curve at negative fields,
correlating it to the presence of the FM$_b$ layer in the NOL SV.
However, the nonsaturation of MR at positive fields is not explained
by our model and had to be explained in terms of irreversible
processes (stable domain walls). The fact that this feature appears
below $\sim$200 K, the same temperature at which exchange and pinned
layer coercive fields showed great enhancement\cite{JOV_Impact}
(thus suggesting the presence of an AFM phase) suggests that it is
related to the magnetic order in the NOL1 layer. The magnetic
structure of the MnIr layer can also have influence in the domain
structure of the FM layer.\cite{Nikitenko}

\acknowledgments{Work supported by projects MMA/1787/95 and
MMA/34155/99 from Fundação Ciência e Tecnologia, Portugal. J. O.
Ventura is thankful for a FCT grant (SFRH/BD/7028/2001).}

\bibliography{Biblio}
\end{document}